\documentclass[american,aps,pra,reprint,superscriptaddress,nofootinbib,floatfix]{revtex4-2}
\usepackage[T1]{fontenc}
\usepackage[utf8]{inputenc}
\usepackage{amsmath}
\usepackage{amssymb}
\usepackage{graphicx}

\makeatletter
\usepackage[T1]{fontenc}
\usepackage[utf8]{inputenc}
\usepackage{amsmath}
\usepackage{amssymb}
\usepackage{bm}
\usepackage{graphicx}

\makeatother

\usepackage{babel}
\begin{document}
\title{Quantum Motion from Local Transition Susceptibility}
\author{Jan Klaers}
\affiliation{Adaptive Quantum Optics, MESA$^{+}$ Institute, University of Twente,
P.O. Box 217, 7500 AE Enschede, The Netherlands}
\begin{abstract}
We characterize quantum motion through the susceptibility of a quantum
state to weak localized state conversion. The susceptibility can be
related to a passage speed characterizing local motion independently
of probability transport. For individual WKB branches, it reproduces
the magnitude of the local dispersion velocity in both propagating
and evanescent regions. In the relativistic theory, the inferred passage
speed is bounded by the speed of light, reaching this bound at the
finite energy corresponding to the center of the mass gap, where the
evanescent decay length equals the reduced Compton wavelength. This
identifies the reduced Compton wavelength as the shortest stationary
evanescent length scale compatible with relativistic quantum motion.
\end{abstract}
\maketitle
Quantum mechanics provides a local conservation law for probability.
The density $\rho$ and probability current $j$ satisfy the continuity
equation, making the current the natural quantity for describing quantum
transport~\cite{Madelung1927,Takabayasi1952}. For a plane wave $\exp(ikx)$,
the ratio $j/\rho=\hbar k/m$ coincides with the group velocity. Beyond
this special case, however, interpreting $j/\rho$ as a local particle
velocity is an additional kinematic assumption rather than a direct
consequence of the Schrödinger equation~\cite{Bohm1952a,Bohm1952b,Holland1993}.
The distinction becomes apparent in stationary states. A standing
wave has $j=0$ because equal counterpropagating branches carry opposite
fluxes and therefore appears to represent a state of vanishing motion.
Nevertheless, it exerts mechanical pressure on a reflecting wall.
Interpreting this pressure in terms of particle scattering requires
a finite incoming momentum current and therefore a nonzero characteristic
speed associated with each branch. Evanescent states raise the same
question. Although they carry no probability flux, they contribute
to mechanical stress. These examples suggest that quantum motion is
not fully characterized by net transport and motivate a definition
that is independent of the probability current.

\section{Transition Susceptibility and Passage Speed}

Our central idea is to probe quantum motion through the response of
a quantum state to a weak localized interaction. We consider a weak
local coupling of rate $J$ acting over a finite interaction region
$\mathcal{C}$ of length $L$ and center $x_{c}$. The coupling transfers
population into an orthogonal state, whose occupation provides the
experimental readout. The directly measurable quantity is the transition
susceptibility,
\begin{equation}
\chi(\mathcal{C},\mathcal{R})\equiv\lim_{J\to0^{+}}\frac{1}{JL}\sqrt{\frac{N_{2}(\mathcal{C},\mathcal{R};J)}{N_{1}(\mathcal{C},\mathcal{R};J)}},\label{eq:susceptibility}
\end{equation}
defined per unit interaction length. Here $N_{1}$ and $N_{2}$ denote
the populations remaining in the initial state and transferred into
the orthogonal state, respectively. Both populations are evaluated
in the comparison region $\mathcal{R}$. The susceptibility is positive
by construction, $\chi(\mathcal{C},\mathcal{R})\ge0$. When the choice
of regions is immaterial, we suppress $\mathcal{C}$ and $\mathcal{R}$
in the notation.

The transition susceptibility is well defined for arbitrary stationary
states, and its inverse has the dimensions of a speed. We therefore
use $\chi^{-1}$ as the natural scale from which a passage speed is
inferred. The relation between the two must be specified by the physical
context and may contain additional dimensionless factors. In the simplest
case, where the response represents a single passage through the interaction
region, this factor can be unity:
\begin{equation}
v_{p}(\mathcal{C},\mathcal{R})=\chi^{-1}(\mathcal{C},\mathcal{R}).\label{eq:passage_speed}
\end{equation}
This simplest relation reflects the idea that particles traversing
the interaction region rapidly experience less state conversion than
particles spending more time inside it. In this work, we first apply
this framework to single nonrelativistic Schrödinger WKB branches.
Coherent superpositions and relativistic states are considered subsequently.

\begin{figure}
\centering{}\includegraphics[width=0.85\columnwidth]{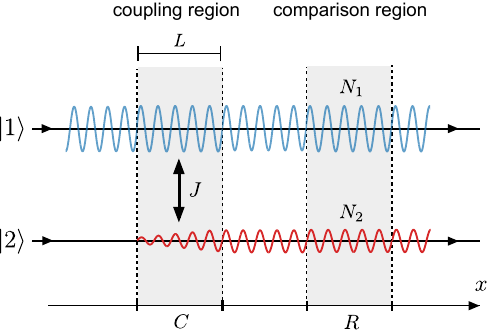}
\caption{Coupling scheme defining the local transition susceptibility. A weak
coupling of rate $J$ acts over the region $\mathcal{C}$ of length
$L$, transferring amplitude from the initial channel $|1\rangle$
to the orthogonal channel $|2\rangle$. The populations $N_{1}$ and
$N_{2}$ are integrated over the same downstream comparison region
$\mathcal{R}$. Their ratio in the weak-coupling limit defines $\chi(\mathcal{C},\mathcal{R})$.
In the simplest single-passage setting, the associated passage speed
is $v_{p}(\mathcal{C},\mathcal{R})=\chi^{-1}(\mathcal{C},\mathcal{R})$;
more generally, the relation between $\chi$ and $v_{p}$ is context
dependent.}
\label{fig:definition} 
\end{figure}

As a minimal realization of the above definition, we consider two
degenerate orthogonal channels, denoted $|1\rangle$ and $|2\rangle$.
The system is prepared in channel $|1\rangle$, while the weak local
coupling is described by $H_{{\rm int}}=\hbar Jw_{\mathcal{C}}(x)\sigma_{x}$,
where $\sigma_{x}$ denotes the Pauli operator acting in the channel
space. The window function $w_{\mathcal{C}}(x)$ is unity inside $\mathcal{C}$
and zero elsewhere. The channel populations $N_{\alpha}(\mathcal{C},\mathcal{R};J)=\int_{\mathcal{R}}|\psi_{\alpha}(x)|^{2}\,dx$
are evaluated in a comparison window $\mathcal{R}$ outside the coupling
region, chosen sufficiently large that small variations of its boundaries
do not affect the population ratio. In the weak-coupling limit, the
leading contribution in channel $\lvert2\rangle$ is linear in $J$.
Consequently, $N_{2}/N_{1}\propto J^{2}$, and $\sqrt{N_{2}/N_{1}}$
gives the magnitude of the linear response. Figure~\ref{fig:definition}
summarizes the measurement geometry.

We now evaluate the transition susceptibility for the elementary case
of a single nonrelativistic Schrödinger WKB branch~\cite{BerryMount1972}.
Here a branch means one of the two local WKB solutions $\psi(x)\simeq Ae^{\pm iqx}$,
where $A$ and $q$ are taken to vary negligibly across the coupling
region. We write the local dispersion as $E=\mathcal{E}(q)$, where
the local wave number $q$ may be real or complex. In the coupling
window it is convenient to work in the basis $|\pm\rangle=(|1\rangle\pm|2\rangle)/\sqrt{2}$,
in which the interaction is diagonal. Since the uncoupled state occupies
only channel $|1\rangle$, it is an equal superposition of the states
$|+\rangle$ and $|-\rangle$. In the stationary eigenvalue problem
the energy $E$ remains fixed, so the weak coupling produces a splitting
of the local wave number rather than of the energy: $\mathcal{E}(q_{\pm})\pm\hbar J=E$.
Here, $q_{\pm}$ denote the wavenumbers associated with the $|\pm\rangle$
states, respectively. Linearizing around the uncoupled wave number
$q$ gives $(\partial\mathcal{E}/\partial q)\delta q_{\pm}\pm\hbar J=0$,
and hence 
\begin{equation}
|q_{+}-q_{-}|=2\hbar J\left|\frac{\partial q}{\partial E}\right|.\label{eq:qsplit}
\end{equation}
Within the coupling window, the stationary solution is of the form
$\left(e^{iq_{+}x}|+\rangle+e^{iq_{-}x}|-\rangle\right)/\sqrt{2}$.
Across a coupling region of length $L$, the channel amplitudes are
\[
\begin{aligned}a_{1} & =\frac{e^{iq_{+}L}+e^{iq_{-}L}}{2}=e^{i\bar{q}L}\cos\!\left(\frac{\vartheta}{2}\right),\\
a_{2} & =\frac{e^{iq_{+}L}-e^{iq_{-}L}}{2}=i\,e^{i\bar{q}L}\sin\!\left(\frac{\vartheta}{2}\right),
\end{aligned}
\]
where we define $\bar{q}=(q_{+}+q_{-})/2$ and $\vartheta=(q_{+}-q_{-})L$.
These expressions remain valid for both real and complex wave numbers.
The common factor $e^{i\bar{q}L}$ cancels in the population ratio,
yielding $N_{2}/N_{1}=|a_{2}/a_{1}|^{2}=|\tan(\vartheta/2)|^{2}$.
In the weak-coupling limit, $|\vartheta|\ll1$, this becomes 
\begin{equation}
\frac{N_{2}}{N_{1}}\simeq\left|\frac{\vartheta}{2}\right|^{2}=\left|\frac{(q_{+}-q_{-})L}{2}\right|^{2}.\label{eq:branch_ratio}
\end{equation}
Combining Eqs.~(\ref{eq:susceptibility}), (\ref{eq:branch_ratio}),
and (\ref{eq:qsplit}) gives $\chi=\hbar|\partial q/\partial E|$.
For this single-passage setting, applying Eq.~(\ref{eq:passage_speed})
therefore gives 
\begin{equation}
v_{p}=\frac{1}{\hbar}\left|\frac{\partial E}{\partial q}\right|.\label{eq:single_branch}
\end{equation}
This reproduces the magnitude of the local dispersion velocity. For
a single branch in a locally uniform region, the factor $L$ cancels
from the susceptibility, and no dependence on $x_{c}$ remains. Consequently,
$\chi$ and $v_{p}=\chi^{-1}$ are constant throughout the region.

For the Schrödinger dispersion relation $E=V+\hbar^{2}q^{2}/(2m)$,
Eq.~(\ref{eq:single_branch}) becomes $v_{p}=\hbar|q|/m$. For a
propagating branch, $q=k$, whereas for an evanescent branch $q=i\kappa$.
Hence, $v_{p}=\hbar k/m$ in the classically allowed region and $v_{p}=\hbar\kappa/m$
in the classically forbidden region. The two cases differ only in
the mechanism driving the state conversion, namely relative phase
accumulation for propagating branches and relative attenuation for
evanescent branches. The evanescent passage speed agrees with earlier
theory~\cite{Klaers2023} and experiment~\cite{Sharoglazova2025},
where its consistency with the dwell time and the Büttiker--Landauer
traversal time is also discussed.

In general, scattering and tunneling theory contain several characteristic
time scales~\cite{Wigner1955,Smith1960,ButtikerLandauer1982,Buttiker1983,Hauge1989,LandauerMartin1994,Winful2006}.
In particular, the position-postselected weak value of momentum~\cite{Aharonov1988,Wiseman2007,DresselJordan2012}
is $p_{{\rm w}}(x)=-i\hbar(\partial_{x}\psi)/\psi$. For a single
local WKB branch, the leading-order approximation gives $p_{{\rm w}}\simeq\pm\hbar q$.
For the nonrelativistic Schrödinger dispersion, this yields $v_{p}=\hbar|q|/m\simeq|p_{{\rm w}}|/m$.
For a general stationary state, however, $|p_{{\rm w}}|/m$ need not
equal the finite-window passage speed defined by the transition susceptibility.

\section{Branch Interference and Mechanical Stress}

We now turn to stationary states composed of more than one branch.
A reflecting wall provides the simplest example. We consider a reflecting
wall at $x=0$ and restrict the motion to the left half-line, $x<0$.
The unperturbed channel-$|1\rangle$ wavefunction is chosen as 
\[
\psi_{1}^{(0)}(x)=e^{ikx}-e^{-ikx}=2i\sin(kx),
\]
with $x<0$. This state satisfies the hard-wall boundary condition
$\psi_{1}^{(0)}(0)=0$. Let the coupling window be $\mathcal{C}=[x_{c}-L/2,x_{c}+L/2]$,
with $x_{c}+L/2<0$. For the Schrödinger Hamiltonian $H_{0}=-\hbar^{2}\partial_{x}^{2}/(2m)$
and $E=\hbar^{2}k^{2}/(2m)$, first-order stationary perturbation
theory gives 
\begin{equation}
(\partial_{x}^{2}+k^{2})\psi_{2}^{(1)}(x)=\frac{2mJ}{\hbar}w_{\mathcal{C}}(x)\psi_{1}^{(0)}(x),\label{eq:wall_first_order}
\end{equation}
where $\psi_{2}^{(1)}=O(J)$ is the first-order correction induced
by the coupling. Let the comparison window lie entirely in the region
between the coupling window and the wall, $\mathcal{R}=[x_{r}-\ell,x_{r}]$
with $x_{c}+L/2<x_{r}-\ell<x_{r}<0$. Imposing $\psi_{2}^{(1)}(0)=0$
and an outgoing-wave condition as $x\to-\infty$, Eq.~(\ref{eq:wall_first_order})
gives $\psi_{2}^{(1)}(x)=C_{2}\sin(kx)$ with
\[
C_{2}=\frac{2JL}{v}\left[1-e^{-2ikx_{c}}\frac{\sin(kL)}{kL}\right],
\]
and $v=\hbar k/m$ throughout the comparison window. Since $\psi_{1}^{(0)}(x)=2i\sin(kx)$
and $\psi_{2}^{(1)}(x)=C_{2}\sin(kx)$ in $\mathcal{R}$, the common
integral over $\sin^{2}(kx)$ cancels, giving 
\[
\frac{N_{2}}{N_{1}}=\left(\frac{JL}{v}\right)^{2}\left|1-e^{-2ikx_{c}}\frac{\sin(kL)}{kL}\right|^{2}.
\]
Using the inverse-susceptibility relation of Eq.~(\ref{eq:passage_speed}),
the corresponding finite-window passage speed becomes 
\begin{equation}
\frac{v_{p}(L,x_{c})}{v}=\left|1-e^{-2ikx_{c}}\frac{\sin(kL)}{kL}\right|^{-1}.\label{eq:left_wall_vp}
\end{equation}
For finite $kL$, the susceptibility depends on both the position
$x_{c}$ and the length $L$ of the coupling window through the coherent
interference of the two counterpropagating branches. For $kL\gg1$,
Eq.~(\ref{eq:left_wall_vp}) approaches $v_{p}(L,x_{c})/v\rightarrow1$.
In this limit, the oscillatory interference term averages out over
the coupling window, and the underlying branch speed is recovered.

\begin{figure}
\centering{}\includegraphics[width=0.93\columnwidth]{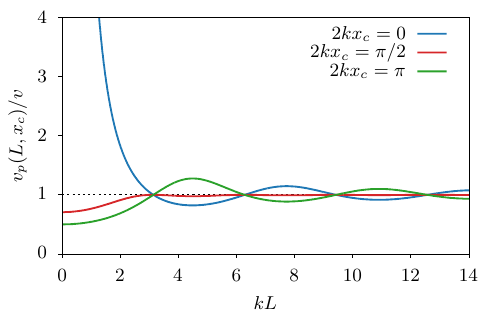}
\caption{Finite-window passage speed for the standing wave formed by reflection
at a hard wall, Eq.~(\ref{eq:left_wall_vp}), shown for three positions
$x_{c}$ of the coupling region. The response depends on $kL$ and
on the standing-wave phase $2kx_{c}\protect\bmod2\pi$ because the
two counterpropagating branches interfere within the coupling region.
For $kL\gg1$, this interference averages out and $v_{p}(L,x_{c})$
approaches the single-branch speed $v=\hbar k/m$. The vertical range
is truncated at $v_{p}/v=4$.}
\label{fig:wall_window}
\end{figure}
Figure~\ref{fig:wall_window} shows this dependence for three positions
of the coupling window. More generally, the relation between the susceptibility
and the inferred passage speed may depend on the location of the comparison
window. In the present geometry, the comparison window is placed between
the coupling window and the reflecting wall. The populations evaluated
there contain the response from a single passage through the coupling
region, so the appropriate relation is $v_{p}=\chi^{-1}$, as used
in Eq.~(\ref{eq:left_wall_vp}). If the comparison window is instead
placed to the left of the coupling window, the susceptibility contains
contributions from two passages. For $kL\gg1$ and a sufficiently
long comparison window, applying the single-passage relation would
give $\chi^{-1}/v\to1/\sqrt{2}$. Accounting for both passages therefore
requires $v_{p}=\sqrt{2}\,\chi^{-1}$, which again recovers the underlying
branch speed $v$. The standing-wave example thus demonstrates that
a finite, nonzero transition susceptibility, and hence a finite passage
speed, can coexist with an identically vanishing net probability current.

The same wall geometry provides a mechanical consistency check of
the passage speed. For a hard wall at $x=0$, the one-dimensional
Schrödinger stress~\cite{NielsenMartin1985,Tokatly2005} exerted
by a state satisfying $\psi(0)=0$ is 
\begin{equation}
F_{{\rm stress}}=\frac{\hbar^{2}}{2m}\left|\partial_{x}\psi(0)\right|^{2}.\label{eq:wall_stress}
\end{equation}
Near the wall, both propagating and evanescent branches can be written
as $\psi(x)=A\left(e^{iqx}-e^{-iqx}\right)$, where $q=k$ for a propagating
branch and $q=i\kappa$ for an evanescent branch. Equation~(\ref{eq:wall_stress})
then gives 
\begin{equation}
F_{{\rm stress}}=\frac{2\hbar^{2}|q|^{2}}{m}|A|^{2}=2m|A|^{2}v_{p}^{2},\label{eq:wall_stress2}
\end{equation}
where we used the previously derived relation $v_{p}=\hbar|q|/m$.
This result can be compared with a particle-scattering picture applied
to the incoming branch $Ae^{iqx}$. Suppose that this branch reaches
the wall with passage speed $v_{p}$ and carries momentum $p$. The
rate at which particles reach the wall is then $|A|^{2}v_{p}$, while
a reflection transfers the momentum $2p$. The corresponding momentum
flux is therefore $F_{{\rm sc}}=2|A|^{2}v_{p}p$. Requiring this momentum
flux to equal the quantum stress Eq. (\ref{eq:wall_stress2}), that
is $F_{\text{stress}}=F_{\text{sc}}$, gives $p=mv_{p}=\hbar|q|$.
The quantum stress is therefore compatible with a particle-scattering
picture governed by the passage speed for both propagating and evanescent
branches.

\section{Relativistic Passage Speed}

In the relativistic setting, inferring passage speed from the measured
susceptibility requires additional care. Previous work relevant to
the evanescent regime includes time-resolved studies of relativistic
tunneling~\cite{Krekora2001,PetrilloJanner2003,Delgado2003}, phase
and dwell times for Dirac and Klein--Gordon barriers~\cite{WinfulRelativistic2004,Bernardini2008},
relativistic Larmor clocks~\cite{KudakaMatsumoto2012}, and quantum-field-theoretic
aspects of causality~\cite{AlkhateebMatzkin2025}. 

We begin by identifying the coupling rate used to normalize the susceptibility
in the relativistic setting. As a specific example, we consider two
weakly coupled waveguides whose longitudinal modes obey an effective
Klein--Gordon equation~\cite{TretyakovAkgun2010}, written in the
standard scalar relativistic form~\cite{Klein1926,FeshbachVillars1958}.
The waveguides extend along the longitudinal $x$ direction and are
separated in the transverse $y$ direction. Their proximity produces
a weak tunnel coupling $J_{y}$ through the overlap of the transverse
modes. Since the susceptibility is defined in the limit $J_{y}\to0$,
the coupling-induced transverse energy splitting is small compared
with $Mc^{2}$, and it is sufficient to retain its leading nonrelativistic
contribution. In the standard modal reduction, the common transverse
mode energy is included in the effective longitudinal mass $M$, while
the remaining transverse dynamics is described by $H_{y}=p_{y}^{2}/(2M)+U(y)$
measured relative to that mode. With $\epsilon=E-V$, the stationary
equation to first order in $H_{y}/(Mc^{2})$ is
\[
\epsilon^{2}\Psi=\left(M^{2}c^{4}+c^{2}p_{x}^{2}+2Mc^{2}H_{y}\right)\Psi.
\]
For two equivalent weakly coupled guides, projection onto the localized
transverse modes~\cite{Yariv1973} gives $H_{y}=\hbar J_{y}\sigma_{x}$.
The coupled longitudinal channels therefore obey
\[
\epsilon^{2}=M^{2}c^{4}+c^{2}p^{2}+2Mc^{2}\hbar J_{y}\sigma_{x},
\]
where $p=p_{x}$ and $J_{y}$ is the microscopic tunnel coupling between
the two guides. In the eigenbasis of $\sigma_{x}$, 
\begin{equation}
\epsilon_{\pm}^{2}=M^{2}c^{4}+c^{2}p_{\pm}^{2}\pm2Mc^{2}\hbar J_{y}.\label{eq:rel_split}
\end{equation}
This notation accommodates both the energy splitting at fixed longitudinal
momentum and the momentum splitting at fixed energy considered below. 

The coupling rate entering the susceptibility is calibrated spectroscopically
at fixed longitudinal momentum $p$. Let $\epsilon_{0}$ denote the
uncoupled energy, $\epsilon_{0}^{2}=M^{2}c^{4}+c^{2}p^{2}$, and write
the coupled branch energies as $\epsilon_{\pm}=\epsilon_{0}+\delta\epsilon_{\pm}$.
Linearizing Eq.~(\ref{eq:rel_split}) using $\epsilon_{\pm}^{2}\simeq\epsilon_{0}^{2}+2\epsilon_{0}\delta\epsilon_{\pm}$
gives the branch shifts $\delta\epsilon_{\pm}=\pm(Mc^{2}/\epsilon_{0})\hbar J_{y}$.
By the energy--angular-frequency relation $\Delta\epsilon=\hbar\Delta\omega$,
the resulting energy splitting corresponds to $\Delta\omega=2J_{{\rm eff}}$,
so that $2\hbar J_{{\rm eff}}=|\epsilon_{+}-\epsilon_{-}|=|\delta\epsilon_{+}-\delta\epsilon_{-}|$.
This yields $\hbar J_{{\rm eff}}=(Mc^{2}/|\epsilon_{0}|)\hbar|J_{y}|$,
or equivalently 
\begin{equation}
J_{{\rm eff}}=\frac{|J_{y}|}{|\gamma_{\epsilon}|},\label{eq:jeff_rel}
\end{equation}
with $\gamma_{\epsilon}=\epsilon_{0}/Mc^{2}$. The linearization requires
$|\delta\epsilon_{\pm}|\ll|\epsilon_{0}|$, or equivalently $\hbar|J_{y}|\ll\gamma_{\epsilon}^{2}Mc^{2}$.
The effective coupling $J_{{\rm eff}}$ is thus fixed by the spectroscopic
energy splitting and will be used to normalize the susceptibility
in both relativistic regimes.

For propagating states ($|\gamma_{\epsilon}|>1$), we write $p_{\pm}=\hbar k_{\pm}$
and define the positive wave-number splitting as $\Delta k=|k_{+}-k_{-}|$.
At fixed energy, $\epsilon_{\pm}=\epsilon_{0}$, while the momenta
are shifted according to $p_{\pm}=p_{0}+\delta p_{\pm}$, where $p_{0}$
is the uncoupled momentum. Linearizing Eq.~(\ref{eq:rel_split})
using $p_{\pm}^{2}\simeq p_{0}^{2}+2p_{0}\delta p_{\pm}$ gives $\delta p_{\pm}=\mp M\hbar J_{y}/p_{0}$
and hence $\Delta k/2=M|J_{y}|/|p_{0}|$. Using $v=|\partial\epsilon_{0}/\partial p_{0}|=c^{2}|p_{0}|/|\epsilon_{0}|=c\sqrt{1-\gamma_{\epsilon}^{-2}}$
together with Eq.~(\ref{eq:jeff_rel}), we obtain 
\begin{equation}
\frac{\Delta k}{2}=\frac{|J_{y}|}{|\gamma_{\epsilon}|v}=\frac{J_{{\rm eff}}}{v}.\label{eq:Delta_k}
\end{equation}
Since Eq.~(\ref{eq:branch_ratio}) remains valid in the relativistic
case, Eq. (\ref{eq:Delta_k}) implies $N_{2}/N_{1}\simeq(J_{{\rm eff}}L/v)^{2}$.
With $J_{{\rm eff}}$ serving as the calibrated coupling in Eq.~(\ref{eq:susceptibility}),
the susceptibility becomes $\chi=1/v$. The relation $v_{p}=\chi^{-1}$
then recovers the usual relativistic particle velocity, $v_{p}=v$,
for $|\gamma_{\epsilon}|>1$.
\begin{figure}
\centering{}\includegraphics[width=1\columnwidth]{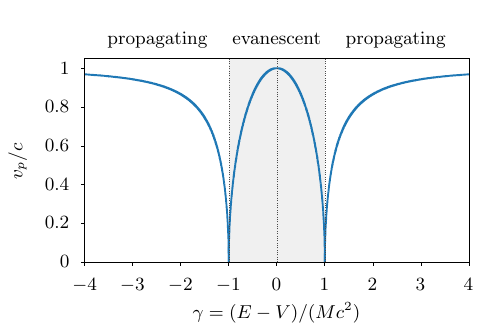}
\caption{Relativistic coordinate passage speed as a function of $\gamma\equiv\gamma_{\epsilon}=(E-V)/(Mc^{2})$.
In the propagating regions, $|\gamma_{\epsilon}|>1$, using $v_{p}=\chi^{-1}$
gives $v_{p}/c=\sqrt{1-\gamma_{\epsilon}^{-2}}$. In the evanescent
mass gap, $|\gamma_{\epsilon}|<1$, compensating for the time dilation
of the reversal process gives $v_{p}/c=\sqrt{1-\gamma_{\epsilon}^{2}}$.
The passage speed is bounded by $c$ and reaches this bound at the
center of the mass gap, where the decay length equals the reduced
Compton wavelength.}
\label{fig:relativistic_branches}
\end{figure}

For evanescent states, the relation between $\chi$ and $v_{p}$ is
less obvious. For a propagating branch, the process admits a direct
semiclassical interpretation: a narrow wave packet moves with the
local dispersion velocity and follows a timelike world line. By contrast,
a stationary evanescent state has no obvious associated world line
because it does not describe directed wave-packet motion. To determine
this relation, we therefore introduce an auxiliary classical stochastic
model that captures the relevant features of stationary evanescent
motion. This model is not part of the susceptibility protocol but
serves to establish the relation between $\chi$ and $v_{p}$ in the
given context. Specifically, we use a two-stream model adapted from
the Goldstein--Kac telegraph process~\cite{Goldstein1951,Kac1974}.
Its trajectories consist of segments with coordinate velocities $+v$
and $-v$, connected by stochastic reversals with direction-dependent
rates. The resulting ensemble reproduces both the zero net transport
and the spatial decay of the evanescent state. The constant segment
speed $v$ represents the local coordinate speed to be inferred from
the susceptibility.

Let $n_{+}(x)$ and $n_{-}(x)$ denote the densities of right- and
left-moving segments. For a stationary evanescent tail, 
\[
n_{+}(x)=n_{-}(x)=\frac{\rho(x)}{2},\qquad\rho(x)\propto e^{-2\kappa x}.
\]
We model the reversal dynamics as a local intrinsic process. Like
spontaneous emission or radioactive decay, the reversal dynamics is
naturally described by rates per unit proper time. For a segment with
coordinate speed $u$, we define the Lorentz factor $\gamma_{v}(u)=1/\sqrt{1-u^{2}/c^{2}}$.
The corresponding laboratory and proper-time rates are related by
\[
\alpha=\gamma_{v}(v)^{-1}\alpha_{\tau},\qquad\beta=\gamma_{v}(v)^{-1}\beta_{\tau},
\]
where $\alpha_{\tau}$ and $\beta_{\tau}$ denote the intrinsic reversal
rates. The stationary two-stream equations become 
\[
\begin{aligned}v\,\partial_{x}n_{+} & =-\alpha n_{+}+\beta n_{-},\\
-v\,\partial_{x}n_{-} & =\alpha n_{+}-\beta n_{-}.
\end{aligned}
\]
Inserting the stationary evanescent density gives 
\begin{equation}
\alpha_{\tau}-\beta_{\tau}=2\kappa v\gamma_{v}(v).\label{eq:reversal_rate}
\end{equation}
Equation~(\ref{eq:reversal_rate}) connects the spatial and temporal
parameters of the process. Compared with the nonrelativistic evanescent
case, it contains the additional factor $\gamma_{v}(v)$, which accounts
for time dilation. The susceptibility measures the weak conversion
response per unit coupling and unit length. Since the proper time
available per unit length is $1/[v\gamma_{v}(v)]$, rather than the
coordinate time $1/v$, the corresponding inverse-susceptibility scale
is $v\gamma_{v}(v)$. Identifying the segment speed with the coordinate
passage speed then gives the self-consistent relation
\begin{equation}
v_{p}=\gamma_{v}(v_{p})^{-1}\chi^{-1},\qquad|\gamma_{\epsilon}|<1.\label{eq:vp_rel}
\end{equation}
Here the Lorentz factor is evaluated at the speed being inferred.

For an uncoupled evanescent state, $p_{0}=i\hbar\kappa_{0}$ and $|\gamma_{\epsilon}|<1$.
The dispersion relation gives $\kappa_{0}=(Mc/\hbar)\sqrt{1-\gamma_{\epsilon}^{2}}$.
At fixed energy, $\epsilon_{\pm}=\epsilon_{0}$, while the decay constants
are shifted according to $\kappa_{\pm}=\kappa_{0}+\delta\kappa_{\pm}$.
Substituting $p_{\pm}=i\hbar\kappa_{\pm}$ into Eq.~(\ref{eq:rel_split})
and linearizing with $\kappa_{\pm}^{2}\simeq\kappa_{0}^{2}+2\kappa_{0}\delta\kappa_{\pm}$
gives $\delta\kappa_{\pm}=\pm MJ_{y}/(\hbar\kappa_{0})$. Defining
the positive decay-constant splitting as $\Delta\kappa=|\kappa_{+}-\kappa_{-}|$,
we therefore obtain $\Delta\kappa/2=|J_{y}|/[c\sqrt{1-\gamma_{\epsilon}^{2}}]$.
Since $q_{\pm}=i\kappa_{\pm}$, Eq.~(\ref{eq:branch_ratio}) gives
the population response $N_{2}/N_{1}\simeq(\Delta\kappa L/2)^{2}$.
Together with Eq.~(\ref{eq:jeff_rel}), the susceptibility definition
gives $\chi=(\Delta\kappa/2)/J_{{\rm eff}}=|\gamma_{\epsilon}|/[c\sqrt{1-\gamma_{\epsilon}^{2}}]$.
If the simplest relation $v_{p}=\chi^{-1}$ were applied without accounting
for the relativistic context, it would yield $v_{p}^{{\rm naive}}=\chi^{-1}=c\sqrt{1-\gamma_{\epsilon}^{2}}/|\gamma_{\epsilon}|$.
This goes to infinity in the center of the mass gap at $\gamma_{\epsilon}=0$.
Using the self-consistent relation in Eq.~(\ref{eq:vp_rel}) gives
\[
v_{p}=\sqrt{1-\frac{v_{p}^{2}}{c^{2}}}\,\frac{c\sqrt{1-\gamma_{\epsilon}^{2}}}{|\gamma_{\epsilon}|}.
\]
Writing $\kappa\equiv\kappa_{0}$ for the decay constant of the uncoupled
evanescent state, the positive solution is
\begin{equation}
v_{p}=c\sqrt{1-\gamma_{\epsilon}^{2}}=\frac{\hbar\kappa}{M},\qquad|\gamma_{\epsilon}|<1.\label{eq:rel_ev_speed}
\end{equation}
Equivalently, the solution satisfies $\gamma_{v}(v_{p})^{-1}=|\gamma_{\epsilon}|$.
Figure~\ref{fig:relativistic_branches} summarizes Eq.~(\ref{eq:rel_ev_speed})
together with the propagating result, showing that the same upper
speed limit applies to propagating and evanescent motion, with saturation
at the center of the mass gap. At $\gamma_{\epsilon}=0$, this result
is understood as the continuous limit $\gamma_{\epsilon}\to0$ taken
after the weak-coupling limit. This order is required by the linearization
condition $\hbar|J_{y}|\ll\gamma_{\epsilon}^{2}Mc^{2}$: for each
nonzero $\gamma_{\epsilon}$, the limit $J_{y}\to0$ is taken first.
Although $\gamma_{v}(v_{p})^{-1}$ vanishes and $\chi^{-1}$ diverges,
their product remains finite and yields $v_{p}=c$.

\section{Conclusion}

We have introduced a characterization of quantum motion based on the
susceptibility of a stationary quantum state to weak local state conversion.
The construction applies directly to stationary states, without reference
to probability current or wave-packet propagation. For a single WKB
branch, the resulting passage speed recovers the magnitude of the
local dispersion velocity. For coherent superpositions, the susceptibility
remains finite even when the net probability current vanishes. In
a standing wave, the inferred passage speed reflects the interference
of the counterpropagating branches and approaches their common branch
speed when this interference is spatially averaged. The same speed
is also compatible with the momentum transfer responsible for wall
pressure.

The same susceptibility definition applies in relativistic Klein--Gordon
theory. In the relativistic theory, the passage process associated
with evanescent states contains both spatial propagation and an intrinsic
reversal dynamics that maintains zero net current. Accounting self-consistently
for the Lorentz time dilation of the reversal dynamics yields the
evanescent passage speed $v_{p}=\hbar\kappa/M\le c$. The corresponding
amplitude decay length can be written directly in terms of the passage
speed,
\[
\ell_{{\rm dec}}\equiv\kappa^{-1}=\frac{\hbar}{Mv_{p}}=\bar{\lambda}_{C}\frac{c}{v_{p}}\ge\bar{\lambda}_{C},\qquad\bar{\lambda}_{C}=\frac{\hbar}{Mc}.
\]
The lower bound is reached when $v_{p}=c$, which occurs at the center
of the mass gap, $\epsilon=0$, corresponding to the finite total
energy $E=V$, despite the finite particle mass. This contrasts with
propagating states, for which the coordinate velocity approaches $c$
only asymptotically as $\lvert\epsilon\rvert\to\infty$. Within the
present framework, the reduced Compton wavelength is therefore the
evanescent decay length at the center of the mass gap and the shortest
stationary decay length compatible with relativistic quantum motion.
This provides an evanescent counterpart to the role of the Compton
scale in relativistic particle localization~\cite{NewtonWigner1949,FoldyWouthuysen1950}.

Taken together, these results connect the passage speed to the magnitude
of the local dispersion velocity, established tunneling-time scales
(see discussion in \cite{Sharoglazova2025}), and the momentum transfer
at a reflecting wall. The relativistic passage speed is bounded by
$c$ and identifies the reduced Compton wavelength as the minimum
evanescent decay length compatible with relativistic quantum motion.
This convergence suggests that susceptibility to local state conversion
captures an aspect of quantum motion that is complementary to probability
transport and remains meaningful even when the probability current
vanishes.
\begin{acknowledgments}
This work received funding from the European Research Council under
the Horizon 2020 research and innovation programme of the European
Union (Grant Agreement No. 101001512).
\end{acknowledgments}

\section*{Data Availability}

No data were created or analyzed in this study. All figures can be
reproduced directly from the equations presented in the article.

\bibliography{passage_speed}

@article{Wigner1955,
  author = {Wigner, E. P.},
  title = {Lower Limit for the Energy Derivative of the Scattering Phase Shift},
  journal = {Phys. Rev.},
  volume = {98},
  pages = {145},
  year = {1955}
}

@article{Smith1960,
  author = {Smith, F. T.},
  title = {Lifetime Matrix in Collision Theory},
  journal = {Phys. Rev.},
  volume = {118},
  pages = {349},
  year = {1960}
}

@article{ButtikerLandauer1982,
  author = {B{\"u}ttiker, M. and Landauer, R.},
  title = {Traversal Time for Tunneling},
  journal = {Phys. Rev. Lett.},
  volume = {49},
  pages = {1739},
  year = {1982}
}

@article{Buttiker1983,
  author = {B{\"u}ttiker, M.},
  title = {{Larmor} Precession and the Traversal Time for Tunneling},
  journal = {Phys. Rev. B},
  volume = {27},
  pages = {6178},
  year = {1983}
}

@article{Hauge1989,
  author = {Hauge, E. H. and St{\o}vneng, J. A.},
  title = {Tunneling Times: A Critical Review},
  journal = {Rev. Mod. Phys.},
  volume = {61},
  pages = {917},
  year = {1989}
}

@article{Aharonov1988,
  author = {Aharonov, Y. and Albert, D. Z. and Vaidman, L.},
  title = {How the Result of a Measurement of a Component of the Spin of a Spin-1/2 Particle Can Turn Out to Be 100},
  journal = {Phys. Rev. Lett.},
  volume = {60},
  pages = {1351},
  year = {1988}
}

@article{Bohm1952a,
  author = {Bohm, D.},
  title = {A Suggested Interpretation of the Quantum Theory in Terms of ``Hidden'' Variables. {I}},
  journal = {Phys. Rev.},
  volume = {85},
  pages = {166},
  year = {1952}
}

@article{Bohm1952b,
  author = {Bohm, D.},
  title = {A Suggested Interpretation of the Quantum Theory in Terms of ``Hidden'' Variables. {II}},
  journal = {Phys. Rev.},
  volume = {85},
  pages = {180},
  year = {1952}
}

@article{Madelung1927,
  author = {Madelung, E.},
  title = {{Quantentheorie in hydrodynamischer Form}},
  journal = {Z. Phys.},
  volume = {40},
  pages = {322--326},
  year = {1927}
}

@article{Takabayasi1952,
  author = {Takabayasi, T.},
  title = {On the Formulation of Quantum Mechanics Associated with Classical Pictures},
  journal = {Prog. Theor. Phys.},
  volume = {8},
  pages = {143--182},
  year = {1952}
}

@book{Holland1993,
  author = {Holland, P. R.},
  title = {The Quantum Theory of Motion: An Account of the {de Broglie--Bohm} Causal Interpretation of Quantum Mechanics},
  publisher = {Cambridge University Press},
  address = {Cambridge},
  year = {1993}
}

@article{Wiseman2007,
  author = {Wiseman, H. M.},
  title = {Grounding {Bohmian} Mechanics in Weak Values and {Bayesianism}},
  journal = {New J. Phys.},
  volume = {9},
  pages = {165},
  year = {2007}
}

@article{DresselJordan2012,
  author = {Dressel, J. and Jordan, A. N.},
  title = {Significance of the Imaginary Part of the Weak Value},
  journal = {Phys. Rev. A},
  volume = {85},
  pages = {012107},
  year = {2012}
}

@article{Yariv1973,
  author = {Yariv, A.},
  title = {Coupled-Mode Theory for Guided-Wave Optics},
  journal = {IEEE J. Quantum Electron.},
  volume = {9},
  pages = {919--933},
  year = {1973}
}

@article{TretyakovAkgun2010,
  author = {Tretyakov, O. A. and Akgun, O.},
  title = {Derivation of {Klein--Gordon} Equation from {Maxwell's} Equations and Study of Relativistic Time-Domain Waveguide Modes},
  journal = {Prog. Electromagn. Res.},
  volume = {105},
  pages = {171--191},
  year = {2010}
}

@article{BerryMount1972,
  author = {Berry, M. V. and Mount, K. E.},
  title = {Semiclassical Approximations in Wave Mechanics},
  journal = {Rep. Prog. Phys.},
  volume = {35},
  pages = {315--397},
  year = {1972}
}

@article{LandauerMartin1994,
  author = {Landauer, R. and Martin, Th.},
  title = {Barrier Interaction Time in Tunneling},
  journal = {Rev. Mod. Phys.},
  volume = {66},
  pages = {217--228},
  year = {1994}
}

@article{Winful2006,
  author = {Winful, H. G.},
  title = {Tunneling Time, the {Hartman} Effect, and Superluminality: A Proposed Resolution of an Old Paradox},
  journal = {Phys. Rep.},
  volume = {436},
  pages = {1--69},
  year = {2006}
}

@article{Klaers2023,
  author = {Klaers, J. and Sharoglazova, V. and Toebes, C.},
  title = {Particle Motion Associated with Wave-Function Density Gradients},
  journal = {Phys. Rev. A},
  volume = {107},
  pages = {052201},
  year = {2023}
}

@article{Sharoglazova2025,
  author = {Sharoglazova, V. and Puplauskis, M. and Mattschas, C. and Toebes, C. and Klaers, J.},
  title = {Energy--Speed Relationship of Quantum Particles Challenges {Bohmian} Mechanics},
  journal = {Nature},
  volume = {643},
  pages = {67--72},
  year = {2025}
}

@article{NielsenMartin1985,
  author = {Nielsen, O. H. and Martin, R. M.},
  title = {Quantum-Mechanical Theory of Stress and Force},
  journal = {Phys. Rev. B},
  volume = {32},
  pages = {3780--3791},
  year = {1985}
}

@article{Tokatly2005,
  author = {Tokatly, I. V.},
  title = {Quantum Many-Body Dynamics in a {Lagrangian} Frame: {I}. {Equations} of Motion and Conservation Laws},
  journal = {Phys. Rev. B},
  volume = {71},
  pages = {165104},
  year = {2005}
}

@article{Klein1926,
  author = {Klein, O.},
  title = {{Quantentheorie und f{\"u}nfdimensionale Relativit{\"a}tstheorie}},
  journal = {Z. Phys.},
  volume = {37},
  pages = {895--906},
  year = {1926}
}

@article{FeshbachVillars1958,
  author = {Feshbach, H. and Villars, F.},
  title = {Elementary Relativistic Wave Mechanics of Spin 0 and Spin 1/2 Particles},
  journal = {Rev. Mod. Phys.},
  volume = {30},
  pages = {24--45},
  year = {1958}
}

@article{Goldstein1951,
  author = {Goldstein, S.},
  title = {On Diffusion by Discontinuous Movements, and on the Telegraph Equation},
  journal = {Q. J. Mech. Appl. Math.},
  volume = {4},
  pages = {129--156},
  year = {1951}
}

@article{Kac1974,
  author = {Kac, M.},
  title = {A Stochastic Model Related to the Telegrapher's Equation},
  journal = {Rocky Mt. J. Math.},
  volume = {4},
  pages = {497--509},
  year = {1974}
}

@article{NewtonWigner1949,
  author = {Newton, T. D. and Wigner, E. P.},
  title = {Localized States for Elementary Systems},
  journal = {Rev. Mod. Phys.},
  volume = {21},
  pages = {400--406},
  year = {1949}
}

@article{FoldyWouthuysen1950,
  author = {Foldy, L. L. and Wouthuysen, S. A.},
  title = {On the {Dirac} Theory of Spin 1/2 Particles and Its Non-Relativistic Limit},
  journal = {Phys. Rev.},
  volume = {78},
  pages = {29--36},
  year = {1950}
}

@article{Krekora2001,
  author = {Krekora, P. and Su, Q. and Grobe, R.},
  title = {Effects of Relativity on the Time-Resolved Tunneling of Electron Wave Packets},
  journal = {Phys. Rev. A},
  volume = {63},
  pages = {032107},
  year = {2001}
}

@article{PetrilloJanner2003,
  author = {Petrillo, V. and Janner, D.},
  title = {Relativistic Analysis of a Wave Packet Interacting with a Quantum-Mechanical Barrier},
  journal = {Phys. Rev. A},
  volume = {67},
  pages = {012110},
  year = {2003}
}

@article{Delgado2003,
  author = {Delgado, F. and Muga, J. G. and Ruschhaupt, A. and Garc{\'i}a-Calder{\'o}n, G. and Villavicencio, J.},
  title = {Tunneling Dynamics in Relativistic and Nonrelativistic Wave Equations},
  journal = {Phys. Rev. A},
  volume = {68},
  pages = {032101},
  year = {2003}
}

@article{WinfulRelativistic2004,
  author = {Winful, H. G. and Ngom, M. and Litchinitser, N. M.},
  title = {Relation between Quantum Tunneling Times for Relativistic Particles},
  journal = {Phys. Rev. A},
  volume = {70},
  pages = {052112},
  year = {2004}
}

@article{Bernardini2008,
  author = {Bernardini, A. E.},
  title = {Delay Time Computation for Relativistic Tunneling Particles},
  journal = {Eur. Phys. J. C},
  volume = {55},
  pages = {125--132},
  year = {2008}
}

@article{KudakaMatsumoto2012,
  author = {Kudaka, S. and Matsumoto, S.},
  title = {{Larmor} Time and Proper Time},
  journal = {Phys. Lett. A},
  volume = {376},
  pages = {3038--3044},
  year = {2012}
}

@article{AlkhateebMatzkin2025,
  author = {Alkhateeb, M. and Matzkin, A.},
  title = {Microcausality and Tunneling Times in Relativistic Quantum Field Theory},
  journal = {Phys. Rev. D},
  volume = {112},
  pages = {076005},
  year = {2025}
}

\end{document}